\documentclass{llncs}

\usepackage[english]{babel}
\usepackage[utf8]{inputenc}
\usepackage{latexsym,amssymb,amstext,amsfonts,amsmath,graphicx,setspace,mathtools,bm}
\usepackage[colorinlistoftodos]{todonotes}
\usepackage{enumerate} 
\usepackage{paralist}
\usepackage{float}
\usepackage{array}
\usepackage{lipsum}

\usepackage{tikz}
\usetikzlibrary{fit,positioning}
\usetikzlibrary{calc}

\newcommand\blfootnote[1]{%
  \begingroup
  \renewcommand\thefootnote{}\footnote{#1}%
  \addtocounter{footnote}{-1}%
  \endgroup
}

\begin{document}

\title{A Continuous Model of Cortical Connectivity}
\titlerunning{Continuous Structural Connectivity}  
%
\author{Daniel Moyer\and Boris A. Gutman \and
Joshua Faskowitz \and Neda Jahanshad\and \\ Paul M. Thompson}
\authorrunning{Daniel Moyer et al.} 
%
\tocauthor{Daniel Moyer, Boris Gutman, Joshua Faskowitz, Neda Jahanshad, and Paul Thompson}
\institute{Imaging Genetics Center, University of Southern California\\
\email{moyerd@usc.edu}}



\maketitle              

\begin{abstract}
We present a continuous model for structural brain connectivity based on the Poisson point process. The model treats each streamline curve in a tractography as an observed event in connectome space, here a product space of cortical white matter boundaries. We approximate the model parameter via kernel density estimation. To deal with the heavy computational burden, we develop a fast parameter estimation method by pre-computing associated Legendre products of the data, leveraging properties of the spherical heat kernel. We show how our approach can be used to assess the quality of cortical parcellations with respect to connectivty. We further present empirical results that suggest the ``discrete'' connectomes derived from our model have substantially higher test-retest reliability compared to standard methods.\blfootnote{Earlier forms of this paper included erroneous results due to a software bug. These errors, which artificially raised ICC scores, have been corrected in this version.}

\end{abstract}

\section{Introduction}

In recent years the study of structural and functional brain connectivity has expanded rapidly.
Following the rise of diffusion and functional MRI
connectomics has unlocked a wealth of knowledge to be explored.
Almost synonymous with the connectome is the network-theory based representation of the brain.
In much of the recent literature the quantitative analysis of connectomes has focused on region-to-region connectivity. This paradigm equates physical brain regions with nodes in a graph, and uses observed structural measurements or functional correlations as a proxy for edge strengths between nodes.

Critical to this representation of connectivity is the delineation of brain regions, the parcellation. Multiple studies have shown that the choice of parcellation influences the graph statistics of both structural and functional networks \cite{satterthwaite2015towards,wang2009parcellation,zalesky2010whole}. It remains an open question which of the proposed parcellations is the optimal representation, or even if such a parcellation exists \cite{de2013parcellation}.


It is thus useful to construct a more general framework for cortical connectivity, one in which \textbf{any} particular parcellation of the cortex may be expressed and its connectivity matrix derived, and one in which the variability of connectivity measures can be modeled and assessed statistically. It is also important that this framework allow comparisons between parcellations, and representations in this framework must be both analytically and computationally tractable. Since several brain parcellations at the macroscopic scale are possible, a representation of connectivity that is independent of parcellation is particularly appealing.


In this paper, we develop such a general framework for a parcellation independent connectivity representation, building on the work of \cite{gutman2014registering}. We describe a continuous point process model for the generation of observed tract\footnote{It is critical to distinguish between white matter fibers (fascicles) and observed ``tracts.'' Here, ``tracts'' denotes the 3d-curves recovered from Diffusion Weighted Imaging via tractography algorithms.} (streamline) intersections with the cortical surface, from which we may recover a distribution of edge strengths for any pair of cortical regions, as measured by the inter-region tract count. Our model is an intensity function over the product space of the cortical surface with itself, assigning to every pair of points on the surface a \emph{point} connectivity. We describe an efficient method to estimate the parameter of the model, as well as a method to recover the region-to-region edge strength. We then demonstrate the estimation of the model on a Test-Retest dataset. We provide reproducibility estimates for our method and the standard direct count methods \cite{jahanshad110alzheimer} for comparison. We also compare the representational power of common cortical parcellations with respect to a variety of measures.

\section{Continuous Connectivity Model}
\label{sec:model}

The key theoretical component of our work is the use of point process theory to describe estimated cortical tract projections. A point process is a random process where any realization consists of a collection of discrete points on a measurable space. 
The most basic of these processes is the Poisson process, in which events occur independently at a specific asymptotic intensity (rate) $\lambda$ over the chosen domain \cite{moller2003statistical}. $\lambda$ completely characterizes each particular process, and is often defined as a function $\lambda: \text{Domain} \rightarrow \mathbb{R}^+$, which allows the process to vary in intensity by location. The expected count of any sub-region (subset) of the domain is its total intensity, the integral of $\lambda$ over the sub-region. In this paper, our domain is the connectivity space of the cortex, the set of all pairs of points on the surface, and the events are estimated tract intersections with the cortical surface.

\subsection{Model Definition}


Let $\Omega$ be union of two disjoint subspaces each diffeomorphic to the 2-sphere representing the white matter boundaries in each hemisphere. Further consider the space $\Omega \times \Omega$, which here represents all possible end point pairs for tracts that reach the white matter boundary. We treat the observation of such tracts as events generated by an inhomogeneous (symmetric) Poisson process on $\Omega \times \Omega$; in our case, for every event $(x,y)$ we have a symmetric event $(y,x)$.



Assuming that each event is independent of all other events except for its symmetric event (i.e., each tract is recovered independently), we model connectivity as a intensity function $\lambda:\Omega \times \Omega \rightarrow \mathbb{R}^+$, such that for any regions $E_1, E_2 \subset \Omega$, the number of events is Poisson distributed with parameter 
\begin{equation}
\mathcal{C}(E_1,E_2) = \iint_{E_1,E_2}{\lambda}(x,y)dxdy.
\end{equation}
Due to properties of the Poisson distribution, the expected number of tracts is exactly $\mathcal{C}(E_1,E_2)$. For any collection of regions $\{E_i\}_{i=1}^N = P$, we can compute a weighted graph $\mathcal{G}(P,\lambda)$ by computing each $\mathcal{C}(E_i,E_j)$ for pairs $(E_i,E_j) \in P \times P$. Each node in this graph represents a region, and each weighted edge represents the pairwise connectivity of the pair of nodes (regions) it connects. We call $P$ a parcellation of $\Omega$ if $\bigcup_i E_i = \Omega$ and $\bigcap_iE_i$ has measure 0 ($\{E_i\}$ is almost disjoint).


\subsection{Recovery of the Intensity Function}

A sufficient statistic for Poisson process models is the intensity function $\lambda(x,y)$. Estimation of the function is non-trivial, and has been the subject of much study in the spatial statistics community \cite{diggle1985kernel}. We choose to use a non-parametric Kernel Density Estimation (KDE) approach due to an efficient closed form for bandwidth estimation described below. This process is self-tuning up to a choice of desiderata for the bandwidth parameter.

We first inflate each surface to a sphere and register them using a spherical registration (See section \ref{sec:pre-proc}); however this entire method can be undertaken without group registration. We treat each hemisphere as disjoint from the other, allowing us to treat $\Omega \times \Omega$ as the product of spheres $(S_1 \cup S_2) \times (S_1 \cup S_2)$. Throughout the rest of the paper $D$ denotes a dataset containing observed tract endpoints $(x,y)_i$, and $\hat{\lambda}$ denotes our estimation of $\lambda$.

The unit normalized spherical heat kernel is a natural choice of kernel for $\mathbb{S}^2$. We use its truncated spherical harmonic representation \cite{chung2006heat}, defined as follows for any two unit vectors $p$ and $q$ on the 2-sphere:
\[K_\sigma(p,q) = \sum_h^H \frac{2h + 1}{4\pi}\exp\{-h(h+1)\sigma\}P^0_h(p\cdot q)\]
Here, $P_h^0$ is the $h^{th}$ degree associated Legendre polynomial of order $0$. Note that the non-zero order polynomials have coefficient zero due to the radial symmetry of the spherical heat kernel \cite{chung2006heat}. However, since we are estimating a function on $\Omega \times \Omega$, we use the product of two heat kernels as our KDE kernel $\kappa$. For any two points $p$ and $q$, the kernel value associated to a end point pair $(x,y)$ is $\kappa((p,q)|(x,y)) = K_\sigma(x,p)K_\sigma(y,q)$. It is easy to show that $\int_{\Omega\times\Omega} K_\sigma(x,p)K_\sigma(y,q)dpdq = 1$.

The spherical heat kernel has a single shape parameter $\sigma$ which corresponds to its bandwidth. While in general tuning this parameter requires the re-estimation of $\hat{\lambda}$ at every iteration, by rewriting our kernel we can memoize part of the computation so that we only need to store the sum of the outer products of the harmonics. Writing out $\kappa((p,q)|D) = \sum_{(x_i,y_i) \in D}K_\sigma(x_i,p)K_\sigma(y_i,q)$, we have the following:
\begin{align*}
\kappa((p,q)|D) 
= & \sum_h^H \sum_k^H \bigg[\underbrace{\left(\frac{2h + 1}{4\pi}\right)\left( \frac{2k + 1}{4\pi} \right)\exp\{-\sigma(h^2+h + k^2+k)\}}_{\text{Independent of } D,~\text{evaluated every iteration}}\\
 & ~~~~~~~~~~~~~~~~~~~~\times \underbrace{\sum_{(x_i,y_i) \in D} P^0_h(x_i\cdot p) P^0_k(y_i\cdot q)}_{\text{Independent of } \sigma,~\text{evaluated } \textbf{once}}\bigg]
\end{align*}
Thus, evaluations of the kernel at any point $(p,q)$ can be done quickly for sequences of values of $\sigma$. We then are left with the choice of loss function. Denoting the true intensity function $\lambda$, the estimated intensity $\hat{\lambda}$, and the leave-one-out estimate $\hat{\lambda}_i$ (leaving out observation $i$), Integrated Squared Error (ISE) is defined:
\begin{align*}
ISE(\sigma | D) = &\int_{\Omega\times\Omega} (\hat{\lambda}(x,y|\sigma) - \lambda(x,y))^2 dxdy\\
\approx & \int\hat{\lambda}(x,y|\sigma)^2 dxdy - \frac{2}{|D|} \sum_{(x_i,y_i) \in D} \hat{\lambda}_{i}(x_i,y_i) + Constant.
\end{align*}
Hall and Marron \cite{hall1987extent} suggest tuning bandwidth parameters using ISE. In practice, we find that replacing each leave-one-out estimate with its logarithm $\log\hat{\lambda}_{i}(x_i,y_i)$ yields more consistent and stable results.

\subsection{Selecting a parcellation}

Given an estimated intensity $\hat\lambda$ and two or more parcellations $P_1,P_2,\dots$, we would like to know which parcellation and associated graph $\mathcal{G}(P,\hat{\lambda})$ best represents the underlying connectivity function. There are at least two perspectives to consider. 

\textbf{Approximation Error}: Because each $P_i$ covers $\Omega$ (and $P_i \times P_i = \Omega \times \Omega$), each $\mathcal{G}(P_1,\hat{\lambda})$ can be viewed as a piece-wise function $g:\Omega \times \Omega \rightarrow \mathbb{R}^+$, where $g(x,y) = \frac{1}{|E_i||E_j|}\mathcal{C}(E_i,E_j)$ such that $x\in E_i$ and $y \in E_j$. In other words, $g$ is the constant approximation to $\lambda$ over every pair of regions. A natural measure of error is another form of Integrated Squared Error: 
\begin{equation}
\label{eq:ISE}
Err(\hat{\lambda},\mathcal{G}(P_1,\hat{\lambda})) = \iint_{\Omega \times \Omega} (g(x,y) - \lambda(x,y))^2 dxdy.
\end{equation} This is analogous to squared loss  ($\ell_2$-loss).

\textbf{Likelihood}: An alternative viewpoint leverages the point process model to measure a likelihood:
\begin{equation}
\label{eq:loglik}
\log\mathcal{L}(P) = \sum_{E_i,E_j \in P} \log\text{Poisson}( |\{(x,y)_i \in D: x \in E_i, y \in E_j\}| ; \mathcal{C}(E_i,E_j)).
\end{equation}
Here, the independence assumption plays a critical role, allowing pairs of regions to be evaluated separately. Unfortunately this is biased toward parcellations with more, smaller regions, as the Poisson distribution has tied variance and mean in one parameter. A popular likelihood-based option that somewhat counterbalances this is Akaike's Information Criterion (AIC), 
\begin{equation}
\label{eq:AIC}
AIC(P) = -2\log\mathcal{L}(P) + |P|(|P|-1).
\end{equation} AIC balances accuracy with parsimony, penalizing overly parameterized models - in our case, parcellations with too many regions.

\section{Application to CoRR Test-Retest Data}

We demonstrate the use of our framework on a test-retest dataset. We measure connectome reproducibility using Intraclass Correlation (ICC) \cite{portney2000statistical}, and compare three parcellations using multiple criteria (See Equations \ref{eq:ISE},\ref{eq:loglik}, and \ref{eq:AIC}).

\subsection{Procedure, Connectome Generation, and Evaluation}

\label{sec:pre-proc}

Our data are comprised of 29 subjects from the Institute of Psychology, Chinese Academy of Sciences sub-dataset of the larger Consortium for Reliability and Reproducibility (CoRR) dataset \cite{zuo2014open}. T1-weighted (T1w) and diffusion weighted (DWI) images were obtained on 3T Siemens TrioTim using an 8-channel head coil and 60 directions. Each subject was scanned twice, roughly two weeks apart. 

T1w images were processed with Freesufer's \cite{fischl2012freesurfer} recon-all pipeline to obtain a triangle mesh of the grey-white matter boundary registered to a shared spherical space \cite{fischl1999high}, as well as corresponding vertex labels per subject for three atlas-based cortical parcellations, the Destrieux atlas \cite{fischl2004automatically}, the Desikan-Killiany (DK) atlas \cite{desikan2006automated}, and the Desikan-Killiany-Tourville (DKT31) atlas \cite{klein2012101}.
Probabilistic streamline tractography was conducted using the DWI in 2mm isotropic MNI 152 space, using Dipy’s \cite{garyfallidis2014dipy} implementation of constrained spherical deconvolution (CSD) \cite{tournier2008resolving} with a harmonic order of 6. 
As per Dipy’s ACT, we retained only tracts longer than 5mm with endpoints in likely grey matter. 


We provide the mean ICC score computed both with and without entries that are zero for all subjects. When estimating $\hat{\lambda}$ the kernels are divided by the number of tracks, and we use a sphere with unit surface area instead of unit radius for ease of computation. We threshold each of the kernel integrated connectomes at $10^{-5}$, which is approximately one half of one unit track density. We then compute three measures of parcellation representation accuracy, namely ISE, Negative Log Likelihood, and AIC scores.

\subsection{Results \& Discussion}
\begin{table*}[h!]
\centering
\begin{tabular}{| c || c | c | c |}
\hline
 Atlas & ~~~~DK~~~~ & ~Destrieux~ & ~~DKT31~~ \\\hline
 Number of Regions & 68 & 148 & 62 \\\hline\hline
 Count ICC & 0.2093 & 0.1722 & 0.2266 \\\hline
 Intensity ICC (Full) & 0.4868 & 0.4535 & 0.4388 \\\hline
 Intensity ICC (w/Threshold) & \textbf{0.5613} & \textbf{0.6481} & \textbf{0.4645} \\\hline
\end{tabular}
\caption{This table shows mean ICC scores for each connectome generation method. The count method - the standard approach - defines edge strength by the fiber endpoint count. The integrated intensity method is our proposed method; in general it returns a dense matrix. However, many of the values are extremely low, and so we include results both with and without thresholding. Highest ICC scores for each atlas are bolded.}
\label{tab:icc}
\end{table*}
\begin{table*}[h!]
\centering
\begin{tabular}{| c || c | c | c |}\hline
 Type & ~~~~DK~~~~ & ~Destrieux~ & ~~DKT31~~ \\\hline
 ISE & $1.8526 \times 10^{-5}$ & $2.1005 \times 10^{-5}$  & $2.1258 \times 10^{-5}$ \\\hline
 Negative LogLik & 85062.5 & 355769.4 & 88444.5 \\\hline
 AIC Score & 174680.95 & 733294.8 & 185253.5 \\\hline
 Retest ISE & $1.0517 \times 10^{-5}$ & $1.0257 \times 10^{-5}$ & $1.1262 \times 10^{-5}$ \\\hline
 Retest Negative LogLik & 85256.0 & 357292.9 & 88434.9 \\\hline
 Retest AIC Score & 175068.1 & 736341.9 & 185234.3 \\\hline
\end{tabular}
\caption{This table shows the means over all subjects of three measures of parcellation ``goodness''. The retest versions are the mean of the measure using the parcellation's regional connectivity matrix (or the count matrix) from one scan, and the estimated intensity function from the other scan.}
\label{tab:comp}
\end{table*}
Table \ref{tab:icc} shows a surprisingly low mean ICC scores for regular count matrices. This may be because ICC normalizes each measure by its $s^2$ statistic, meaning that entries in the adjacency matrices that should be zero but that are subject to a small amount of noise -- a few erroneous tracks -- have very low ICC. Our method in effect smooths tracts endpoints into a density; end points near the region boundaries are in effect shared with the adjacent regions. Thus, even without thresholding we dampen noise effects as measured by ICC. With thresholding, our method's performance is further improved, handily beating the counting method with respect to ICC score. It is important to note that for many graph statistics, changing graph topology can greatly affect the measured value \cite{zalesky2010whole}. While it is important to have consistent non-zero measurements, the difference between zero and small but non-zero in the graph context is also non-trivial. The consistency of zero-valued measurements is thus very important in connectomics.

Table \ref{tab:comp} suggests that all three measures, while clearly different, are consistent in their selection at least with respect to these three parcellations. It is somewhat surprising that the Destrieux atlas has quite low likelihood criteria, but this may be due to the (quadratically) larger number of region pairs. Both likelihood based retest statistics also choose the DK parcellation, while ISE chooses the Destrieux parcellation by a small margin. It should be noted that these results must be conditioned on the use of a probabilistic CSD tractography model. Different models may lead to different intensity functions and resulting matrices. The biases and merits the different models and methods (e.g. gray matter dilation for fiber counting vs streamline projection) remain important open questions. 


\begin{figure*}[t]
\includegraphics[width=\textwidth]{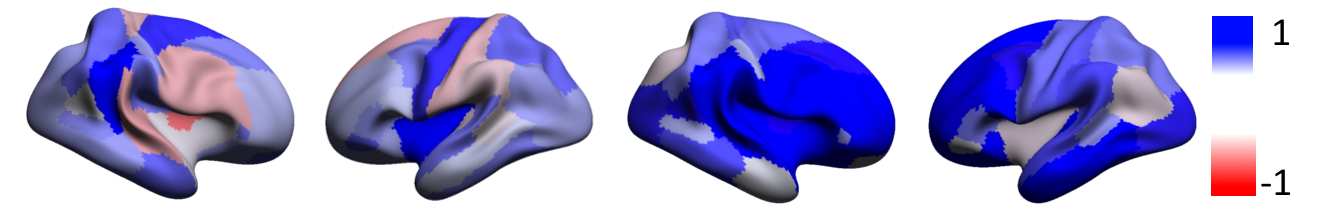}
\caption{A visualization of the ICC scores for connectivity to Brodmann Area 45 (Destrieux region 14) for the Count connectomes (\textbf{left}) and the proposed Integrated Intensity connectomes (\textbf{right}). {\color{blue}Blue} denotes a higher score.}
\label{fig:icc}
\end{figure*}

\begin{figure*}[t]
\includegraphics[width=\textwidth]{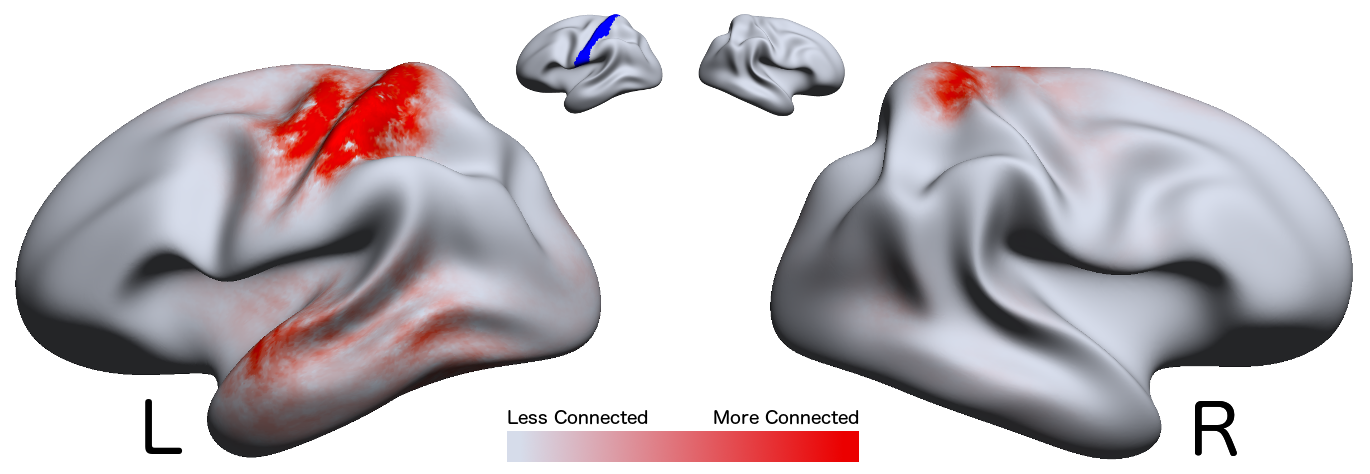}
\caption{A visualization of the marginal connectivity $M(x) = \int_{E_i} \hat{\lambda}(x,y) dy$ for the Left Post-central Gyrus region of the DK atlas (Region 57). The region is shown in {\color{blue} blue} on the inset. {\color{red} Red} denotes higher connectivity regions with the {\color{blue} blue} region.}
\label{fig:r57}
\end{figure*}

\section{Conclusion}
\vspace{-0.35cm}
We have presented a general framework for structural brain connectivity. This framework provides a representation for cortical connectivity that is independent of the choice of regions, and thus may be used to compare the accuracy of a given set of regions' connectivity matrix. We provide one possible estimation method for this representation, leveraging spherical harmonics for fast parameter estimation. We have demonstrated this framework's viability, as well as provided a preliminary comparison of regions using several measures of accuracy.

The results presented here lead us to conjecture that our connectome estimates are more reliable compared to standard fiber counting, though we stress that a much larger study is required for strong conclusions to be made. Further adaptations of our method are possible, such as using FA-weighted fiber counting. Our future work will explore these options, conduct tests on larger datasets, and investigate the relative differences between tracking methods and parcellations more rigorously.

\section*{Acknowledgments}

This work was supported by NIH Grant U54 EB020403, as well as the Rose Hills Fellowship at the University of Southern California. The authors would like to thank the reviewers as well as Greg Ver Steeg for multiple helpful conversations. 

\bibliographystyle{splncs03}
\bibliography{concon.bib}

\end{document}